\documentstyle[emulateapj,psfig]{article}
\def\ale{\mathrel{\hbox{\rlap{\hbox{\lower4pt\hbox{$\sim$}}}\hbox{$<$}}}}
\def\age{\mathrel{\hbox{\rlap{\hbox{\lower4pt\hbox{$\sim$}}}\hbox{$>$}}}}

\def\arcmin{\hbox{$^\prime$}}
\def\arcsec{\hbox{$^{\prime\prime}$}}
\def\fd{\hbox{$~\!\!^{\rm d}$}}
\def\fh{\hbox{$~\!\!^{\rm h}$}}
\def\fm{\hbox{$~\!\!^{\rm m}$}}
\def\fs{\hbox{$~\!\!^{\rm s}$}}

\def\farcs{\hbox{$.\!\!^{\prime\prime}$}}         
\def\gsim{\mathrel{\hbox{\rlap{\lower.55ex \hbox {$\sim$}}
                   \kern-.3em \raise.4ex \hbox{$>$}}}}
\def\lsim{\mathrel{\hbox{\rlap{\lower.55ex \hbox {$\sim$}}
                   \kern-.3em \raise.4ex \hbox{$<$}}}}

\def\aj#1#2#3{\bibitem[]{}#1, AJ, #2, #3.}
\def\apj#1#2#3{\bibitem[]{}#1, {\it Ap. J.}, {\bf#2}, #3.}
\def\apjlett#1#2#3{\bibitem[]{}#1, {\it Ap. J. (Letters)}, {\bf #2},
#3.}

\def\iauc#1#2{\bibitem[]{}#1, IAU Circ.~No.~#2}
\def\gcn#1#2{\bibitem[]{}#1, GCN Circ.~No.~#2}

\def\nature#1#2#3{\bibitem[]{}#1, {\it Nature}, {\bf #2}, #3.}

\def\pasp#1#2#3{\bibitem[]{}#1, PASP, #2, #3.}

\lefthead{Bloom et al.}
\righthead{Transient Discovery of GRB 980703}
\begin{document}

\submitted{Submitted to {\it ApJ Letters} on 27 August 1998}
 
\title{\large \bf The Discovery and Broad-band follow-up of \\
the Transient Afterglow of GRB 980703}

\author{J.~S.~Bloom\altaffilmark{1},
  D.~A.~Frail\altaffilmark{2}, S.~R.~Kulkarni\altaffilmark{1},
  S.~G.~Djorgovski\altaffilmark{1}, J.~P.~Halpern\altaffilmark{3},
  R.~O.~Marzke\altaffilmark{4}, D.~R.~Patton\altaffilmark{5},
  J.~B.~Oke\altaffilmark{6,1}, K.~D.~Horne\altaffilmark{7},
  R.~Gomer\altaffilmark{8}, R.~Goodrich\altaffilmark{9},
  R.~Campbell\altaffilmark{9},
  G.~H.~Moriarty-Schieven\altaffilmark{10},
  R.~O.~Redman\altaffilmark{6}, P.A.~Feldman\altaffilmark{6},
  E.~Costa\altaffilmark{11}, N.~Masetti\altaffilmark{12}}

\altaffiltext{1}{California Institute of Technology, Palomar Observatory,
105-24, Pasadena, CA 91125, USA}

\altaffiltext{2}{National Radio Astronomy Observatory, P.~O.~Box O,
  Socorro, NM 87801, USA}

\altaffiltext{3}{Columbia University, Columbia Astrophysics
  Laboratory, 538 W.~120th St., New York, NY 1002, USA}

\altaffiltext{4}{Observatories of the Carnegie Institution of
Washington, 813 Santa Barbara St., Pasadena, CA 91101, USA}

\altaffiltext{5}{Department of Physics and Astronomy, University of
Victoria, P.O.~Box 3055, Victoria, BC, V8W 3P6, Canada}

\altaffiltext{6}{Herzberg Institute of Astrophysics, National Research
  Council, 5071 West Saanich Road, Victoria, BC, V8X 4M6, Canada}

\altaffiltext{7}{University of St.~Andrews, School of Physics and
Astronomy, North Haugh, KY16 9SS St.~Andrews, Scotland}

\altaffiltext{8}{Howard Hughes Medical Institute, Dept.~of
  Biochemistry and Cell Biology, MS-140, Rice University, Houston, TX
  77005-1892, USA}

\altaffiltext{9}{W.~M.~Keck Observatory, 65-0120 Mamalahoa Highway,
Kamuela, HI 96743, USA}

\altaffiltext{10}{Joint Astronomy Centre, 600 A'Ohoku Place, Hilo, HI
  96720, USA}

\altaffiltext{11}{Istituto di Astrofisica Spaziale, CNR, via Fosso del 
Cavaliere, Roma I-00133, Italy}

\altaffiltext{12}{Istituto Tecnologie e Studio Radiazioni
Extraterrestri, CNR, Via Gobetti 101, Bologna I-40129, Italy}

\begin{abstract}

  We report on the discovery of the radio, infrared and optical
  transient coincident with an X-ray transient proposed to be the
  afterglow of GRB 980703.  At later times when the transient has
  faded below detection, we see an underlying galaxy with $R=22.6$;
  this galaxy is the brightest host galaxy (by nearly 2 magnitudes) of
  any cosmological GRB thus far.  In keeping with an established
  trend, the GRB is not significantly offset from the host galaxy.
  Interpreting the multi-wavelength data in the framework of the
  popular fireball model requires that the synchrotron cooling break
  was between the optical and X-ray bands on July 8.5 UT and that the
  intrinsic extinction of the transient is $A_{\rm V}=0.9$.  This is
  somewhat higher than the extinction for the galaxy as a whole, as
  estimated from spectroscopy.

\end{abstract}

\section{Introduction}

The gamma-ray burst (GRB) 980703 of 1998 July 3.18 UT was detected by
the All Sky-Monitor (ASM) of the Rossi X-ray Timing Explorer, Ulysses,
BATSE, and the Gamma-Ray Burst Monitor of BeppoSAX (Levine, Morgan, \&
Muno 1998; Hurley \& Kouveliotou 1998; Kippen 1998; Amati et
al.~1998). The 4\arcmin-radius ASM position was further refined to a
50\arcsec-radius error circle with observations using the Narrow Field
Instruments on BeppoSAX (Galama et al.~1998a).

Radio observations were begun on 1998 July 4.41 UT at the Very Large
Array\footnotemark\footnotetext{The VLA is a facility of the National
  Science Foundation operated under cooperative agreement by
  Associated Universities, Inc.} (VLA) in Socorro, NM. A single, weak
radio source was seen coincident with an optically variable source
(Frail et al.~1998a).  Subsequent
optical\footnotemark\footnotetext{Some optical and infrared
  observations were obtained at the W.~M.~Keck Observatory which is
  operated by the California Association for Research in Astronomy, a
  scientific partnership between the California Institute of
  Technology, the University of California and the National
  Aeronautics and Space Administration.} and radio observations
(Zapatero Osorio et al.~1998; Vreeswijk et al.~1998; Bloom et
al.~1998a; Frail et al.~1998b) confirmed the transient nature of the
source. Following a now-familiar pattern, we identify the radio and
optical transient to be the afterglow of GRB 980703.

\section{Observations}

We present a log of the optical and infrared (IR) imaging photometry
in Table \ref{tab:obs}. Data were obtained using a CCD imager at the
Palomar 60-inch telescope, and LRIS (Oke et al.~1995) and NIRC
(Matthews \& Soifer 1994) instruments at the Keck 10-m telescopes. All
nights were photometric.  Flux calibration of the optical images was
performed in the Johnson-Kron-Cousins magnitude system using the
standard field PG\thinspace{2213$-$006} (Landolt 1992), including a
color term and an atmospheric extinction correction.  The IR data were
calibrated using the standards SJ 9186 and SJ 9101 (Persson 1998). In
Figure 1 we give a finding chart for the transient.

VLA monitoring of GRB 980703 was carried out at 1.43, 4.86 and 8.46
GHz. In addition, an observation was made on 1998 July 10.53 UT using
the SCUBA array on the James Clerk Maxwell
Telescope\footnotemark\footnotetext{The James Clerk Maxwell Telescope
  is operated by The Joint Astronomy Centre on behalf of the Particle
  Physics and Astronomy Research Council of the United Kingdom, the
  Netherlands Organization for Scientific Research, and the National
  Research Council of Canada.} (JCMT) at 220 GHz. At 1.43 GHz the
source has remained weak ($<120$ $\mu$Jy). At 4.86 GHz the source
exhibited a large degree of variability, consistent with interstellar
scattering. At 8.46 GHz 
\noindent

\begin{minipage}[h]{8.7cm}
\centerline{\psfig{file=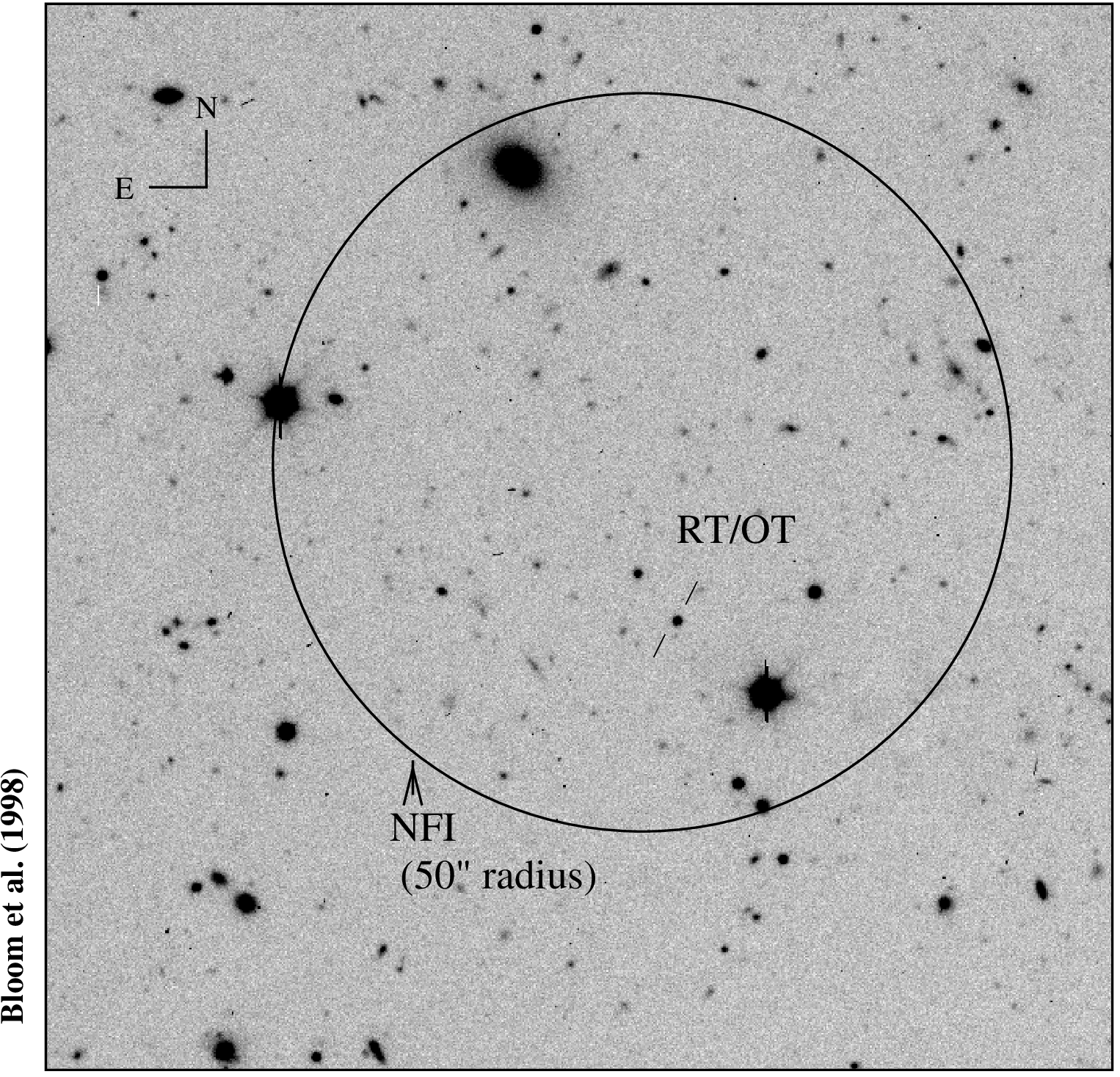,angle=270,width=8.65cm}} 
\figcaption{Finding chart for the
transient afterglow of GRB 980703. The radio/optical transient
(``RT/OT'') was discovered in the overlapping region of the ASM
position (Smith, Levine, \& Muno 1998), IPN annulus (Hurley \&
Kouveliotou 1998), and BeppoSAX NFI position (Galama et al.~1998a).
The radio transient is located at
\hbox{$\alpha$: 23\fh 59\fm 6\fs .666 $\pm{~0\farcs 05}$},
\hbox{$\delta$: +08\fd 35\arcmin 7\farcs 07 $\pm{~0\farcs 05}$}
(J2000). The RT/OT position is marked within the 3\arcmin $\times$
3\arcmin\ LRIS 600-s R-band image taken on July 6.6 (see Table
\ref{tab:obs})
.}\label{fig:dis}
\end{minipage}
\vspace{0.3cm}

\noindent the flux has been relatively steady with a
mean of 940 $\mu$Jy during the first 11 days after the GRB. Despite
excellent conditions, no 220 GHz source was visible above a 2-$\sigma$
error limit of 5.2 mJy. The observed scintillation and the broad-band
spectrum of GRB 980703 are similar to other well-studied radio
afterglows (e.g.~Shepherd et al.~1998, Taylor et al.~1998).  Further
discussion on the radio properties of this GRB can be found in Frail
et al.~(1998c).

\section{Results}

In Table \ref{tab:obs} we summarize the measured fluxes at the
position of the transient.  In Figure 2 we plot the
optical and IR light-curves of the transient source. It is a
prediction of spherical fireball models that the late-time light
curves of GRB afterglow should closely follow a power-law decline and,
indeed, this has been observed (e.g.~Galama et al.~1998b; Zharikov,
Sokolov, \& Baryshev 1998).  The presence of a host galaxy will cause
the light-curve to flatten eventually. We thus model the light-curve
as $f_{\rm total} = f_{0}t^{\alpha} + f_{\rm host}$, with the decay
constant ($\alpha$), and the corresponding normalization of the
transient flux ($m_0$), and the host galaxy flux ($m_{\rm host}$).  In
Table \ref{tab:fit} we give our light-curve fits to this OT + host
flux model. Only the $R$- and $I$-band had a sufficient number of data
points to fit for the three parameters independently. The weighted
average of decay constants for the $I$-band and $R$-band (Table
\ref{tab:fit}) give $\alpha = -1.17 \pm{0.25}$.  For the IR bands, we
assume $\alpha = -1.17$ and fit for $m_{\rm host}$ and $m_0$. In the
case of the $B$-band and $V$-band we report the last observed
magnitude as $m_{\rm host}$.

It is of interest to quantify any potential offset of the transient
from its host galaxy as the distribution of offsets may help to
constrain the various progenitor models. Using the methodology
outlined in Bloom et al.~(1998b), we registered the Keck LRIS images
taken on July 6 and July 18 (see Table \ref{tab:obs}) accounting for
the relative scale, rotation, translation, and second-order
distortion.  The r.m.s.~centroid differences of the transformed star
positions (including both axes) was $\sigma = 36$ milliarcsec.  We
find the angular separation of the OT and the host galaxy to be 63
milliarcsec, which includes the error of the transformation and
centering errors of the objects themselves.  However, the transient
contributed only 30\% of the total OT + host flux on the first epoch.
Thus, we conclude the transient is offset by 210 $\pm$ 120
milliarcsec; this angular separation is consistent with no offset from
the host galaxy at the 1.8-$\sigma$ level.

\section{Discussion and conclusions}
\begin{figure*}[tb]
 \centerline{\psfig{file=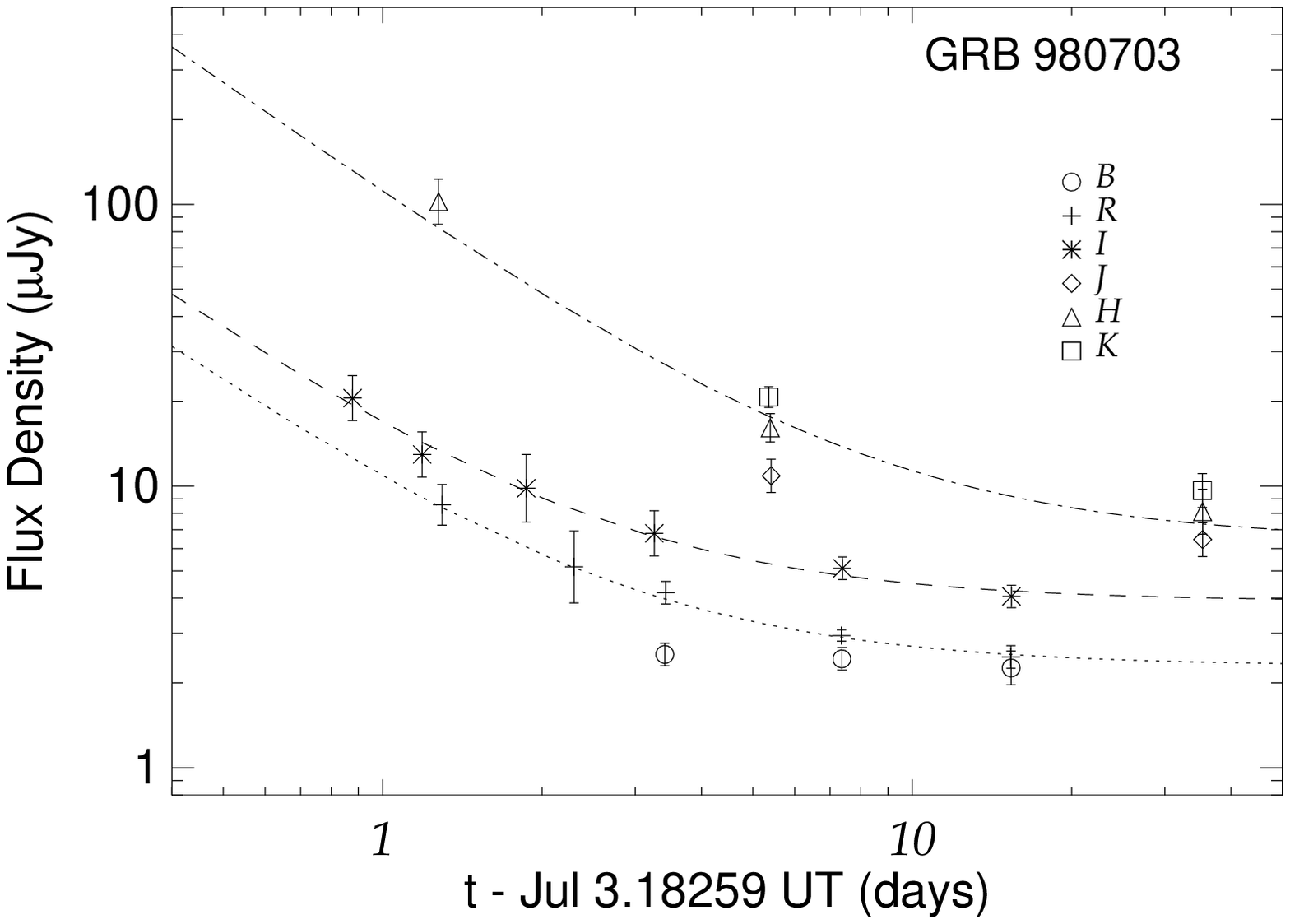,height=3.2in,width=3.0in}
    \psfig{file=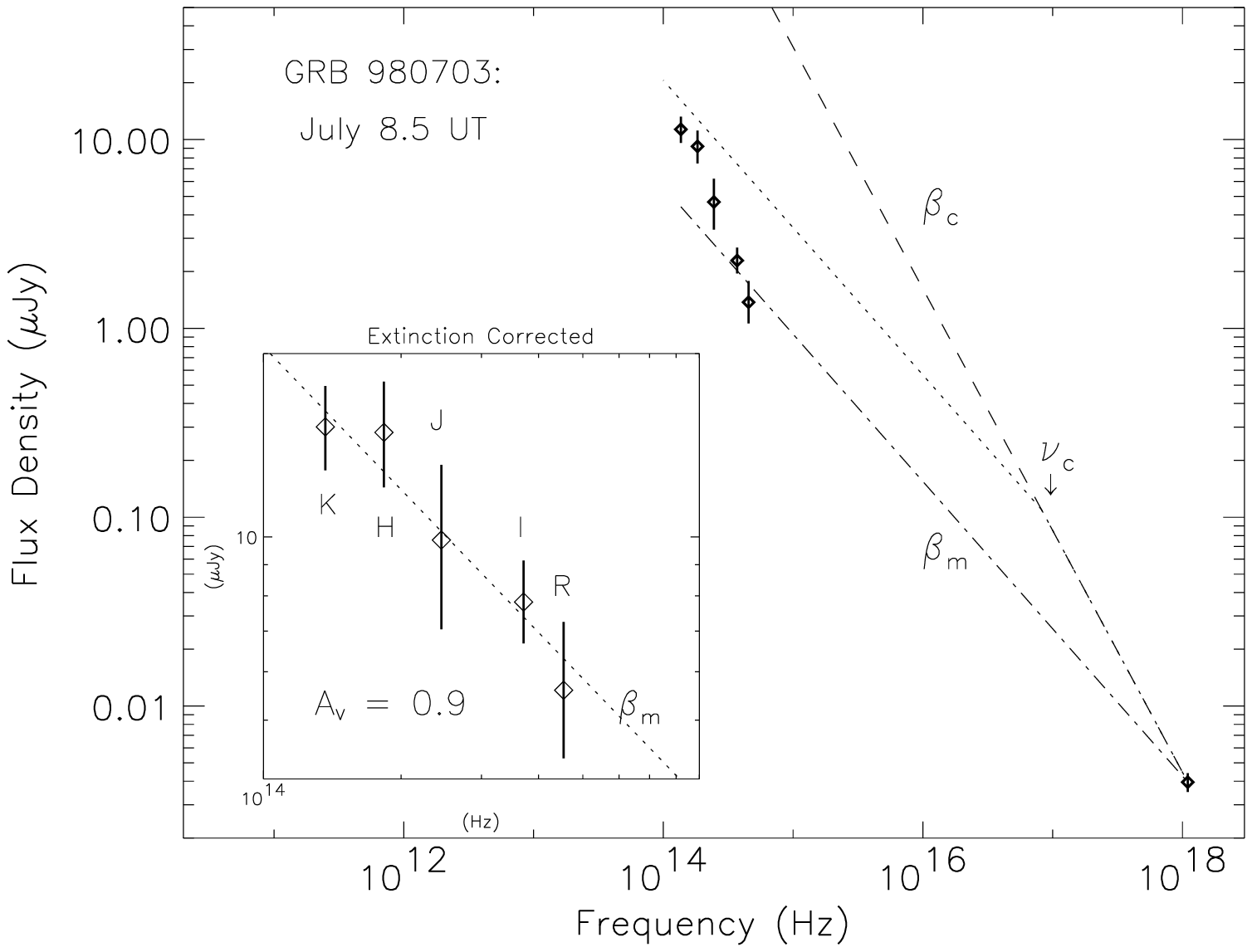,height=3.2in,width=3.5in}}
\caption[Light Curve of 980703]{\footnotesize (left) Optical and
infrared light curve of the transient of GRB 980703.  The fit to the
$R$-, $I$-, and $H$-band light curves are shown (see Table
\ref{tab:fit}).  The first 4 $I$-band points are from Vreeswijk et
al.~(1998), and the first $H$-band point from Henden et
al.~(1998). The plotted fluxes have not been corrected for Galactic
extinction. We did not fit a model to the $B$-band light-curve since
the OT had faded away at $B$-band by the time of our first
observation. (right) Inferred broad-band spectrum spectrum of the
transient afterglow of GRB 980703 on July 8.5.  The fluxes plotted
above have been corrected for Galactic extinction (see Table
\ref{tab:fit}).  We show the predicted spectra of the IR/optical/X-ray
if these frequencies were simultaneously in the synchrotron phase
($F_{\nu} \propto \nu^{\beta_m}$; dot-dashed line) or in the
synchrotron cooling regime ($F_{\nu} \propto \nu^{\beta_c}$; dashed
line).}
\label{fig:ltcurve}
\end{figure*} 
The small offset of the afterglow from the underlying galaxy suggests,
like most other well-studied GRBs (e.g.~Sahu et al.~1997; Odewahn et
al.~1998), that GRB 980703 is closely connected to its host galaxy.
If indeed the galaxy is the host, then at $R = 22.6$ it is the
brightest (apparent) host galaxy of any cosmological GRB thus far (see
Hogg \& Fruchter 1998 for a review).  In a companion paper Djorgovski
et al.~(1998) derive a redshift for the OT + host of $z=0.966$.
Assuming $H_0 = 65$ km s$^{-1}$ Mpc$^{-1}$ and $\Omega = 0.2$, the
luminosity distance to the host is $d_L = 1.92 \times 10^{28}$ cm.  A
cubic-spline fit to the host spectrum (Table \ref{tab:fit}) gives a
flux density of $4.6 \mu$Jy at 8745 \AA, the redshifted effective
wavelength of the $B$-band.  This corresponds to an absolute magnitude
of $M_B \approx -21.2$.  The host is thus a typical galaxy at the knee
of the $B$-band luminosity function at the epoch of $z \simeq 1$
(e.g.~Lilly et al.~1995).  In the July 18 LRIS images the galaxy
appears marginally extended with a FWHM of 0.74\arcsec\ in
0.65\arcsec\ seeing.

In discussing the interpretation of the specific features of the
optical-IR spectrum we will assume an adiabatic hydrodynamical
evolution of the afterglow and that the cooling frequency ($\nu_c$) is
blue-ward of the peak frequency ($\nu_m$).  See Sari, Piran, \&
Narayan (1998) for details.

The optical and IR transient flux at day 5.3 (July 8.5 UT) are fit by
a spectral slope of $\approx -2$ with some indication of a break near
1 $\mu$m (Figure 2). In the case of GRB 971214, where
the behavior was quite similar, the IR/optical data required an
additional extinction correction (presumably dust extinction local to
the GRB) to fit the expectations of the fireball model (Ramaprakash et
al.~1998; Wijers \& Galama~1998).  Indeed, without an extinction
correction, 
an extrapolation of the IR/optical spectrum would
under-predict the X-ray point (Galama et al.~1998a) by several orders
of magnitude.  If the IR/optical/X-ray behavior were simultaneously in
the cooling regime, the spectral slope expected from our derived
$\alpha$ would be $\beta_c = -1.1$.  No value of extinction can
correct the observed IR/optical spectral slope without {\it under-predicting} the X-ray point by more than a factor of ten. We therefore
exclude the possibility that the IR/optical/X-ray were simultaneously
in the cooling regime on July 8.5.  The other possibility is that
$\nu_c$ had not yet moved into the IR/optical bands; in this case we
expect a spectral index of $\beta_m = -0.78$.  Indeed, fitting the
IR/optical data with a galactic extinction curve to this expected
power-law slope, we find that with an extinction correction of $A_{\rm
V} = 0.9\pm{0.2}$ (Castro-Tirado et al.~1998 found $A_{\rm V}
\approx 1.6$), the IR/optical data can be made consistent with $\beta_m =
-0.78$.  Nevertheless, extrapolating the extinction-corrected spectrum
{\it over-predicts} the X-ray flux by a factor of 10 (see Figure 2).

To this end, we suggest that there must be a cooling break ($\nu_c
\simeq 5 \times 10^{16}$ Hz) between the IR/optical and X-ray bands on
July 8.5 UT.  This observation leads immediately to the value for
electron energy spectral index $p = 2.6$, consistent with other
afterglows (e.g.~Yoshida et al.~1998; Sokolov et al.~1998).
Furthermore, the measured decay constants are consistent with this
picture. Since the X-ray transient was observed while in the cooling
regime its time decay ($\alpha_{\rm X-ray} = -1.33 \pm {0.25}$; Galama
et al.~1998a) is {\it expected} to be steeper than that measured from
the IR/optical ($\alpha = -1.17 \pm 0.25$).

Our model fits to early-time observations do allow us to constrain
additional properties of the afterglow. Assuming the standard bandpass
shapes and zero-magnitude calibration (Fukugita, Shimasaku, \&
Ichikawa 1995; Bessel \& Brett 1988), our fit (Table \ref{tab:fit}) to
the transient light-curve, and an extinction of $A_{\rm V} = 0.9$, we
find a total integrated fluence over $R$-, $I$-, $J$-, $H$-, and
$K$-bands from day 1 onwards of $S_{\rm IR/opt} \approx 8 \times
10^{-8}$ erg cm$^{-2}$. For comparison, the fluence above 20 keV in
the GRB itself was $S_{\gamma} \approx 4.6 \times 10^{-5}$ erg
cm$^{-2}$ (Kippen 1998).  We estimate the total X-ray fluence (1.3 ---
10 keV) from day one onwards of $S_{\rm X-ray} \approx 6 \times
10^{-7}$ erg cm$^{-2}$ using the data reported in Galama et
al.~(1998a). The fluence ratio is then $S_{\rm IR/opt}$: $S_{\rm
X-ray}$ : $S_{\gamma}$ = 1 : 8 : 575.  The associated total energy
release ($\simeq 4\pi S d_L^2/(1+z)$) is $E_\gamma = 10^{53}$ erg,
$E_{\rm X-ray} = 10^{51}$ erg, and $E_{\rm IR/opt} = 2 \times 10^{50}$
erg. Since $E_\gamma$ is much larger than the sum of the inferred
energy release at all other frequencies (including radio), we suggest
the shock cannot be radiative, but rather is adiabatic.

In conclusion, the broad-band spectrum and the light curves
demonstrates the general consistency of the afterglow spectrum and the
expectations of an adiabatic relativistic blastwave.  Correction of
the intrinsic extinction suggests that the synchrotron cooling break
frequency was somewhere between the optical and X-ray bands on July
8.5 UT.  Since the inferred intrinsic extinction is $\approx 0.6$ mag
greater than the extinction inferred toward the host's HII regions
(Djorgovski et al.~1998), this suggests that the GRB could have
occurred in a dusty environment.

\acknowledgements

We thank F.~Chaffee, the Director of the W.~M.~Keck Observatory, for
allocation of service time observations.  The staff of WMKO and of the
Palomar Observatory are thanked for their continued effort on the GRB
project. We thank W.~Wack, T.~Williams, E.~Morris, M.~Pierce, and
A.~Conrad for service observations at Keck and P.~Groot for assistance
during our latest Keck run. In addition, we are indebted to the
BeppoSAX team and the RXTE-ASM team for their X-ray work.  The GRB
community, as a whole, has greatly benefited from the information
disseminated by the GRB Coordinates Network (GCN), and we thank, in
particular S.~Barthelmy for his efforts.  SRK's and JSB's research is
supported by NSF and NASA.  SGD acknowledges a partial support from
the Bressler Foundation.

\begin{deluxetable}{lccccccc}
\footnotesize
\tablecaption{Photometric Observations of GRB 980703
\label{tab:obs}} \tablewidth{0in} \tablehead{ \colhead{Date} &
\colhead{Instr.} & \colhead{Band} & \colhead{Int.} & \colhead{Seeing}
& \colhead{Mag.\tablenotemark{a}} & \colhead{Err.}
\\ \colhead{UT} & \colhead{} & \colhead{} &
\colhead{(s)} & \colhead{(arcsec)} & \colhead{} & \colhead{}& \colhead{}
}\tablecolumns{7} \startdata
Jul 4.48 & P60& R & 600 & 0.85  & 21.28 & 0.18   
\nl
Jul 5.482 & P60& R & 600 & 1.9 & 21.83 & 0.32 
\nl
Jul 6.607 & LRIS & R & 600 & 0.65 &  22.06 & 0.1  
\nl
Jul 6.607 & LRIS & B & 600 & 0.65 &  23.05 & 0.1  
 \nl 
Jul 8.576 & NIRC & H & 50  & 0.45 &  19.56 & 0.13  
\nl 
Jul 8.545 & NIRC & K & 100 & 0.45 &  18.78 & 0.10  
\nl
Jul 8.597 & NIRC & J & 50  & 0.45 &  20.45 & 0.15  
\nl
Jul 10.55 & LRIS & B & 750 & 1.0  &  23.09 & 0.1   
\nl
Jul 10.56 & LRIS & V & 600 & 1.0  &  22.94 & 0.1   
\nl
Jul 10.53 & LRIS & R & 600 & 1.0  &  22.44 & 0.1  
\nl
Jul 10.57 & LRIS & I & 120 & 1.0  & 21.61 & 0.1  
\nl
Jul 18.57 & LRIS & B & 600 & 0.5 & 23.17 & 0.15   
\nl
Jul 18.57 & LRIS & V & 300 & 0.5 &  23.01 & 0.1   
\nl
Jul 18.55 & LRIS & R & 600 & 0.5 &  22.63 & 0.1  
\nl
Jul 18.58 & LRIS & I & 150 & 0.5 &  21.86 & 0.1  
\nl
Aug~7.53 & NIRC & J & 120 & 0.5 & 21.02 & 0.15  
\nl
Aug~7.52 & NIRC & H & 120 & 0.5 & 20.30 & 0.2 
\nl
Aug~7.50 & NIRC & K & 120 & 0.5 & 19.61 & 0.15  
\enddata
\tablenotetext{a}{Optical data are reported in the
Johnson-Kron-Cousins magnitude system and are uncorrected for
Galactic extinction. The reported magnitudes are corrected (by $ <
0.02$ mag) using the standard atmospheric transmission at the
respective observing sites.}
\end{deluxetable}

\begin{deluxetable}{cccc}
\footnotesize
\tablecaption{Broad-band Spectrum Fits\tablenotemark{\dagger}
\label{tab:fit}} \tablewidth{0in}
\tablehead{ \colhead{Band} 
& \colhead{$\alpha$ } & \colhead{$m_0$} &
\colhead{$m_{\rm host}$}}\tablecolumns{6} \startdata 
$B$ & \ldots & \ldots & $22.92\pm{0.2}$ \nl
 $V$ & \ldots & \ldots & $22.83\pm{0.1}$ \nl
 $R$& $-1.22\pm{0.35}$ & $21.14^{+0.23}_{-0.29}$ & $22.62^{+0.18}_{-0.22}$ \nl
 $I$  &$-1.12\pm{0.35}$ & $20.51^{+0.15}_{-0.17}$ &$21.92^{+0.22}_{-0.28}$ \nl
 $J$  & (-1.17) & $19.16_{-0.50}^{+0.34}$ & $21.11^{+0.16}_{-0.19}$\nl
 $H$  & (-1.17) & $17.72_{-0.20}^{+0.17}$ & $20.66^{+0.24}_{-0.31}$\nl
 $K$  & (-1.17) & $17.20^{+0.19}_{-0.23}$ & $19.78^{+0.17}_{-0.20}$\nl
\enddata 
\tablenotetext{\dagger}{Magnitudes are corrected assuming a
Galactic extinction of $E(B-V) = 0.0607$ (Schlegel, Finkbeiner, \&
Davis 1998) and the extinction curve from Cardelli, Clayton, \& Mathis
(1989). The uncertainties on a given parameter are formal errors of
the fit and include the uncertainties in the other parameters. }
\end{deluxetable}


\begin{thebibliography}{99}

\gcn{Amati, L., Frontera, F., Costa, E., \& Feroci, M.~1998}{146}

\pasp{Bessel, M.~S., \& Brett, J.~M.~1988}{100}{1134}

\gcn{Bloom, J.~S., Djorgovski, S.~G., Kulkarni, S.~R., \& Frail,
D.~A.~1998a}{136}

\bibitem[]{}Bloom, J.~S., Djorgovski, S.~G., Kulkarni, S.~R., \& Frail,
D.~A.~1998b, {\it Ap.~J.~(Letters)}, accepted.

\apj{Cardelli, J. A., Clayton, G. C., \& Mathis, J. S.~1998}{345}{245}

\bibitem[]{}Castro-Tirado, A. J. et al. 1998, ApJ, submitted.

\bibitem[]{}Djorgovski, S.~G., Kulkarni, S.~R., Bloom, J.~S.,
  Goodrich, R., Frail, D.~A., Piro, L., \& Palazzi, E.~1998, ApJ
  Lett., in press

\gcn{Frail, D.~A., Halpern, J.~P., Bloom, J.~S., Kulkarni, S.~R.,
Djorgovski, S.~G.~1998a}{128}

\gcn{Frail, D.~A., Kulkarni, S.~R., Bloom, J.~S., \& Djorgovski,
S.~G.~1998b}{141}

\bibitem[]{}Frail, D.~A.~et al.~1998c, in preparation.

\pasp{Fukugita, M., Shimasaku, K., \& Ichikawa, T.~1995}
{107}{945}

\gcn{Galama, T.~J.~et al.~1998a}{127}

\apjlett{Galama, T.~J. et al.~1998b}{497}{13}

\bibitem[]{} Hogg, D. W., \& Fruchter, A. S. 1998, astro-ph/9807262.

\gcn{Kippen, R.~M.~1998}{143}

\gcn{Hurley, K., \& Kouveliotou, C.~1998}{125} 

\aj{Landolt, A.~1992}{104}{34} 

\iauc{Levine, A., Morgan, E., \& Muno, M.~1998}{6966} 

\apj{Lilly, S.~J., Tresse, L., Hammer, F., Crampton, D., Le F\'evre,
O.~1995}{455}{108}

\bibitem[]{}{Matthews, K., \& Soifer, B.~T.~1994, Infrared Astronomy
with Arrays, the Next Generation, ed.~I.~McLean (Dordrecht: Kluwer),
239.}

\bibitem[]{}Odewahn, S.~C.~et al., {\it Ap.~J.~(Letters)}, in press
(astro-ph/9807212).

\pasp{Oke, J.~B., et al.~1995}{107}{375}

\bibitem[]{}\hbox{Persson, S.~E.~1998} 
\hbox{({\tt http://astro.caltech.edu/mirror/}}
\hbox{ {\tt keck/realpublic/inst/nirc/HST\_Standards.html} ).}

\nature{Ramaprakash, A.~N.~et al.~1998}{393}{43}

\nature{Sahu, K.~C., et al.~1997}{387}{476}

\apjlett{Sari, R., Piran, T., \& Narayan, R.~1998}{497}{L17}

\apj{Schlegel, D. J., Finkbeiner, D. P., \& Davis, M.~1998}{500}{525}

\apj{Shepherd, D.~S., Frail, D.~A., Kulkarni, S.~R., \& Metzger,
M.~R.~1998}{497}{859}

\gcn{Sokolov, V.~V., Zharikov, S.~V., \& Panthenko, I.~1998}{147}

\apjlett{Taylor, G.~B.~et al.~1998}{502}{115}

\gcn{Vreeswijk, P.~M.~et al.~1998}{132}

\bibitem[]{}Wijers, R.~A.~M.~J., \& Galama, T.~J.~1998, {\it Ap.~J.}, submitted.

\bibitem[]{}Yoshida et al.~1998, in 4th Huntsville GRB Sym.,
eds.~C.~Meegan, R.~Preece and T.~Koshut, (New York: AIP), 441.

\gcn{Zapatero Osorio, M.~R., Castro-Tirado, A.,
Gorosabel, J., Oscoz, A., Kemp, S., Frontera, F., \& Nicastro,
L.~1998}{130}

\bibitem[]{}Zharikov, S.~V., Sokolov, V.~V., \& Baryshev, Y.~V.~1998, A\& A, 337, 356.
\end{thebibliography}
\end{document}